\documentclass[referee]{aa}
\usepackage{graphicx,natbib}
\usepackage[latin1]{inputenc}
\usepackage[T1]{fontenc}
\bibpunct{(}{)}{;}{a}{}{,}
\usepackage{txfonts}
\usepackage{changebar}
\usepackage{bm}
\usepackage{units}
\usepackage{color}

\def\mearth{M_{\oplus}}

\def\f1{f_{\rm I}}

\def\beq{\begin{equation}}
\def\eeq{\end{equation}}

\def\mdot2D{\dot M_{\rm 2D}}
\def\mdot3D{\dot M_{\rm 3D}}

\def\mearth{M_\oplus}

\def\mearth{M_\oplus}

\def\simgr{\,\hbox{\hbox{$ > $}\kern -0.8em \lower 1.0ex\hbox{$\sim$}}\,}
\def\simle{\,\hbox{\hbox{$ < $}\kern -0.8em \lower 1.0ex\hbox{$\sim$}}\,}
\def\beq{\begin{equation}}
\def\eeq{\end{equation}}

\def\simgr{\,\hbox{\hbox{$ > $}\kern -0.8em \lower 1.0ex\hbox{$\sim$}}\,}
\def\simle{\,\hbox{\hbox{$ < $}\kern -0.8em \lower 1.0ex\hbox{$\sim$}}\,}
\def\beq{\begin{equation}}
\def\eeq{\end{equation}}

\def\apj{ApJ}                 
\def\apjl{ApJ}                
\def\apjs{ApJS}               
\def\ao{Appl.Optics}          
\def\aap{A\&A}                
\def\mnras{MNRAS}             

\def\({\left(}
\def\){\right)}
\def\<{\left<}
\def\>{\right>}

\def\({\left(} 
\def\){\right)} 
\def\<{\left<} 
\def\>{\right>} 

\def\bc{\begin{changebar}}
\def\bce{\begin{center}}
\def\beq{\begin{equation}} 

\def\bi{\begin{itemize}}

\def\btab{\begin{tabular}{p{1.7cm}p{12cm}p{1.5cm}}}
\def\bt2{\begin{tabular}{p{1 cm}p{4.5cm}p{10cm}}}

\def\ec{\end{changebar}}
\def\ece{\end{center}}
\def\eeq{\end{equation}} 
\def\ei{\end{itemize}}

\def\etab{\end{tabular}\\}

\def\mH2{m_\mathrm{H_2}}

\def\dmax0{\rho_\mathrm{max}}
\def\dmaxS0{\Sigma_\mathrm{max}}

\def\rH2{r_\mathrm{H_2}}

\def\r0max{r_\mathrm{0max}}

\def\s0{\sigma_\mathrm{0}}

\def\xp0{x_{\rm{M0}}}

\def\z0max{z_\mathrm{0max}}

\def\apj{{\it ApJ}}                 
\def\apjl{ApJ}                
\def\apjs{ApJS}               
\def\ao{Appl.Optics}          
\def\aap{{\it A\&A}}                
 
\def\mnras{MNRAS}             

\begin{document}

\title{The maximum mass of planetary embryos formed in core-accretion models}

\author{Y. Alibert \inst{1}}
\offprints{Y. Alibert}
\institute{Physikalisches Institut  \& Center for Space and Habitability, Universitaet Bern, CH-3012 Bern, Switzerland, 
        \email{yann.alibert@space.unibe.ch}
          }

\abstract
{In the core-accretion model, the typical size of solids accreted to form planetary embryos and planetary cores is debated. First models assumed that the major
part of planetary cores came from large size planetesimals, but other more recent models are based on the accretion of low size pebbles.}
{The goal of this paper is to compute the maximum mass a growing planetary embryo can reach depending on the size of accreted planetesimals or pebbles,
and to infer the possibility of growing the cores of giant planets, and giant planets themselves.}
{We compute the internal structure of {the gas envelope of} planetary embryos, to determine the core mass that is necessary to bind an envelope large enough to
destroy planetesimals or pebbles while they are gravitationally captured. We also consider the effect of the advection wind originating from 
the protoplanetary disk, following the results of Ormel et al. (2015).}
{We show that for low mass pebbles, once the planetary embryo is larger than a fraction of the Earth mass, the envelope is large enough to destroy
and vaporize pebbles completely before they can reach the core. The material constituting pebbles is therefore released in the planetary envelope,
and later on dispersed in the protoplanetary disk, if the advection wind is strong enough. As a consequence the growth of the planetary embryo is stopped
at a mass that is so small that Kelvin-Helmholtz accretion cannot lead to the accretion of significant amounts of gas.
For larger planetesimals, a similar process occurs but at much larger mass, of the order of ten Earth masses, and is followed by rapid
accretion of gas.}
{{If the effect of the advection wind is as efficient as described in Ormet al. (2015), }the combined effect of the vaporization 
of accreted solids in the envelope of forming planetary embryos, and of this advection wind, prevents the growth of the planets 
at masses smaller or similar to the Earth mass in the case of formation by pebble accretion, {up to a distance of the order of 10 AU. In the case of
formation by accretion of large mass planetesimals, the growth of the planetary core is limited at masses of the order of ten Earth masses.
However, contrary to the case of pebble accretion, further growth is still possible and proceeds either via the accretion of gas, or via the accretion of solids destroyed in the 
planetary envelope once the effect of the advection wind has ceased, when the planetary Hill radius becomes comparable to the disk scale height.}}


\maketitle

\section{Introduction}
\label{sec:introduction}

In the core-accretion model, planetary embryos grow from the accretion of solids, up to the point they are massive
enough to start accrete noticeable amounts of gas (Pollack et al. 1996, Ida and Lin 2004, Alibert et al. 2005, Alibert et al. 2013, Benz et al. 2014)
in a runaway process. However, before this phase of runaway accretion, planetary embryos already start to gravitationally bind
a small envelope when they are rather small (smaller than the Earth). Even if the envelope  is at this stage tiny,
and therefore does not contribute noticeably to the total planetary mass, it is important for the planetary growth.
Indeed,  solids present in the protoplanetary disk interact with this tiny envelope, via gas drag and heating, and this enlarges the cross-section
of the planet and therefore the accretion rate of solids {(see e.g. Podolak et al., 1988, Pollack et al., 1996, Inaba and Ikoma, 2003, Alibert et al., 2005)}. 
The effect of the envelope depends strongly on the mass and density of accreted solids, and, generally speaking, is larger for smaller 
accreted bodies (we will see below, however, that this general rule has some exceptions).

One point that is currently very debated, in the context of the core-accretion model, is the typical size and mass of accreted
bodies. In the first models, the most important mass carrier were planetesimals, whose typical size can be intermediate (of the order
of kilometer, see Ida and Lin 2004, Fortier et al. 2013), or large (of the order of one hundred of kilometer, see Pollack et al. 1996,
Alibert et al. 2005).  One problem of these models is that the growth of the planetary core can be slow, especially at large 
distances from the star, with an accretion rate that is typically of the order of $10^{-6} M_\oplus / $yr (see e.g. Alibert et al. 2013).
{Note that even if core accretion begins with planetesimals larger than one kilometer, collisional fragments
can contribute greatly to core formation (Inaba et al., 2003, Kobayashi et al., 2010, Kobayashi et al. 2011).}

Recently, it has been proposed that the main component (in term of total accreted mass) of captured solids could be much smaller
ones, called pebbles, whose typical size is of the order of centimeter. These pebbles are therefore strongly coupled to
the gas, as their Stockes number (the product of the Keplerian frequency and the stopping time due to gas drag) is of the
order of 0.1 (e.g. Ormel \& Klahr, 2010, Lambrechts \& Johansen, 2012, 2014, Bitsch et al. 2015, Levison et al. 2015). In this case, the accretion cross-section
of planetary embryos is strongly increased due to the presence of the gas envelope, and the growth of planetary cores can be very
rapid. The  accretion rate for embryos around one Earth mass is typically one order of magnitude larger than in the
case of planetesimals, with rates of the order of $10^{-5} M_\oplus / $yr (see e.g. Lambrechts et al. 2014).

The presence of the planetary envelope, however, does not only increase the cross section of planetary embryos, it also
strongly affect accreted solids. Indeed, incoming solids suffer gas drag, heating from the ambient gas and from the large
post-shock temperature (in case of supersonic trajectories), and, in some case, mechanical destruction (when the difference
between the pressure in front of the incoming body and the one on the sides is larger than the tensile strength of the body).
If the gas envelope is hot  and massive enough, incoming solids are destroyed, and their material vaporized and
lost in the gas envelope before they can reach the core (Podolak et al. 1988, Mordasini, Alibert, Benz, 2006). 
Beyond this point, the growth of the planetary core stops, but not the growth of the whole planet, as accreted solids increase 
the metal content of the planetary envelope. In addition, the end of core growth coincides with a strong decrease of the core's luminosity,
which triggers  accretion of H/He gas from the protoplanetary disk on a Kelvin-Helmholtz (hereafter KH) timescale 
{(Ikoma et al., 2000, Hubickyj et al., 2005)}. As the metal and H/He content 
of the envelope increase, the planet can eventually become super-critical and runaway H/He accretion occurs (see Venturini et al. 
2016 for formation calculations taking into account self-consistently the enrichment of the planetary envelope).

The previous framework is based on the fact that the planetary envelope is a closed system that practically can only gain mass (solids
or H/He) from the protoplanetary disk. However, recent calculations show that this may not always be the case. Indeed, using 
3D hydrodynamical models, Ormel et al. (2015) have shown that the envelope of the growing planets is constantly replenished
by gas coming from the protoplanetary disk, on a timescale that depends on the planetary and disk properties. When this effect 
is dominant the growth of a planet beyond the point where solids are destroyed in the envelope suffers two problems. First, the 
Kelvin-Helmholtz contraction can be hindered (if the KH timescale is longer than the replenishment timescale, see 
Ormel et al. 2015). Second, the material lost by solids in the envelope may be lost to the protoplanetary disk, if the deposition rate 
of such material is smaller than the loss rate by the advection wind. This occurs if the pollution timescale (the time on which the 
composition of the envelope is noticeably changed) is longer than the replenishment timescale. The first effect prevents the 
increase of the HHe component of the envelope, the second prevents the increase of the metal content of the envelope. Altogether, 
the mass growth of the planet is stopped. Therefore, if the replenishment is efficient enough, the destruction of accreted solids 
results  in the end of the planetary growth, and therefore in the existence of a maximum mass of planetary embryos. 
Any further growth must  proceed either by collision of such planetary embryos, or must wait until the conditions have changed (e.g. 
the replenishment timescale, which depends on the properties of the protioplanetary disk, becomes larger).

\section{Theoretical model}
\label{model}

\subsection{Planetary envelope}
\label{envelope}

We compute the planetary structure by solving the planetary internal structure equations, assuming the luminosity is only given
by the accretion rate of solids, $L = G M_{\rm core} \dot M_{\rm core} / R_{\rm core}$. This means that we \textit{de facto} assume  
that the solids reach the solid core. This is true until the point where solids are destroyed in the envelope (beyond this point,
we do not use anymore the internal structure equations).

\begin{eqnarray}
{d r^3 \over d m} & = & {3 \over 4 \pi \rho}, \\
{d P \over d m}&  = & {- G ( m + M_{\rm core} ) \over 4 \pi r^4} , \\
{d T \over d P}&  = & min(\nabla_{\rm conv}, \nabla_{\rm rad}),
\end{eqnarray}

In these equations, $r,P,T$ are respectively the radius, the pressure
and the temperature inside the {envelope}. These three quantities depend on the
gas mass $m$ {between the surface of the core and the sphere of radius $r$}, the distance in the planetary envelope
towards the planetary center. $\rho$ is the mass density
given as a function of T and P by the EOS of Saumon et al. (1995),
and $M_{\rm core}$ the mass of the solid core.
The temperature gradient is given by either the radiative gradient ($\nabla_{\rm rad}$):
\begin{equation}
\nabla_{\rm rad} = {{ 3 \kappa L } \over { 64 \pi \sigma G ({m + M_{\rm core}}) T^3}}
\end{equation}
or the convective gradient, equal to the adiabatic one. In these formulas, 
 $\sigma$ is the Stefan-Boltzman constant, $G$ the gravitational constant.
Finally, the luminosity $L$ which enters in the computation of the radiative gradient
is given by the accretion energy of solids, and the opacity $\kappa$ used here is taken to be equal to the 
one of interstellar medium (Bell and Lin 1994). Indeed, when the planetary envelope is small enough
so that accreted solids are not destroyed (the phase during which we use the internal structure equations), 
the planetary envelope is constantly replenished by gas coming from the protoplanetary disk, and pollution 
by accreted solids is negligible. The population of grains in the planet is therefore close to the one in the 
protoplanetary disk, which is assumed to be similar to the one in the interstellar medium.
 
These equations are solved, using as boundary conditions the pressure and temperature in the protoplanetary 
disk at the position of the planetary embryo and defining the planetary radius as a combination of the Hill and Bondi 
radius (Lissauer et al., 2009):
\begin{equation}
R_{\rm planet} = {G M_{\rm planet} \over { \left( C_{\rm S}^2 + 4 G M_{\rm planet} / R_{\rm Hill} \right) } }
\end{equation}
where $C_{\rm S}^2 $ is the square of the sound velocity in the protoplanetary disk at the planet's location $a_{\rm planet}$,
and $R_{\rm Hill} = a_{\rm planet} \left( { M_{\rm planet} \over 3 M_\odot } \right)$.

We note that, although the envelope is replenished on a timescale that can be rather short, we can still use the equations
quoted above to determine the internal structure of the envelope. Indeed, the internal structure equations are valid for timescales
long compared to the dynamical timescale, which is much shorter than the replenishment timescale. In addition, the equation giving the
temperature gradient is valid only when the planetary envelope is optically thick. For the opacity we consider (Bell and Lin 1994),
this happens for a planetary core equal to a fraction of an Earth mass (see also Bodenheimer and Pollack, 1986).

\subsection{Protoplanetary disk model}

The thermodynamical properties of the disk are computed using the model of Bitsch et al. (2015). {This model is based on fits
of 2D   radiative transfer calculations including the effect of the irradiation from the star. The choice of the disk model is
however not very important for the results we present in this paper (the disk model is more important when dealing for example
with planetary migration which depends strongly on the local disk structure).} This models allows the determination
of the temperature structure in the protoplanetary disk as a function of distance to the star and accretion rate in the disk. The accretion 
rate in the disk evolves with time on a 1 Myr timescale {(Hartmann et al. 1998)} .
\begin{equation}
\log {M_{\rm dot} \over {M_\odot / yr}} = -8.00 -1.40  \log { \left( t \over {\rm 1 Myr} \right) }
\end{equation}

From the value of the accretion rate $M_{\rm dot} $, we compute the temperature in the solar nebula using the formulas presented in 
Bitsch et al. (2015). The surface density and the other properties of the protoplanetary disk (scale height, mid plane density) are computed 
once the value of the turbulent parameter $\alpha_{\rm SS}$ is chosen. Our reference value is $\alpha_{\rm SS} = 10^{-3}$, this value does
not have a strong influence on our results.

\subsection{Timescales}
\label{timescales}

Four timescales are important in the problem considered here. The first one is the replenishment timescale,
that we compute using the formulas of Ormel et al. (2015):
\begin{equation}
t_{\rm replenish} = { M_{\rm enve} \over f^*_{\rm cover} R_{\rm Bondi}^2 \left( R_{\rm Bondi} \Omega \right) \rho_{\rm disc} }
\end{equation}
where $R_{\rm Bondi}$ is the Bondi radius of the planetary embryo, $\Omega$ the Keplerian frequency, $M_{\rm enve}$ the mass of
the planetary envelope computed using the method presented in Sect. \ref{envelope}, and $ f^*_{\rm cover}$ is a numerical parameter
that quantifies the fraction of the planetary envelope that is directly part of streamlines coming from the protoplanetary disk and is found from
numerical simulations to be in the range [0.1 - 1] (see Ormel et al. 2015). We use here a value of 0.1, which leads to an upper boundary of the likely
value of $t_{\rm replenish}$.

The simulations of Ormel et al. (2015) have been performed in the case of an embedded planet, where $R_{\rm Bondi}$ is smaller than
the scale-height of the disc $H_{\rm disc}$. In our simulations, we therefore consider that the replenishment process stops when the planet
is no more embedded, namely when $H_{\rm disc} < R_{\rm Bondi}$.

The second important timescale is the Kelvin-Helmoltz timescale, which governs the growth of the planet once the supply of energy source
at the planetary core has ceased (this energy comes from the accretion of pebbles or planetesimals). Lee et al. (2014) have shown
that, for dust-free envelopes, the Kelvin-Helmoltz timescale is given by 
\begin{equation}
t_{\rm KH, \chi = 0.5} =  10^6 \left( {Z \over 0.02} \right)^{0.25}  \left( { M_{\rm core} \over 5 M_\oplus } \right) ^{-3.93} {\rm yr}
\end{equation}
This scaling is derived for a gas-to-core ratio $\chi$ of 0.5 {and is based on calculations taking into account
the gas opacity, but no dust opacity (see Lee et al. 2014)}. {We note that this timescale is  longer than the one derived by Hori and Ikoma (2010).
We come back in Sect. \ref{evolution} on the consequences of a much reduced KH timescale.}

In the case of dusty envelopes ({assuming ISM like dust grains, see Lee et al. 2014}), the KH timescale scales with
$Z^{0.72}$ and the pre-factor is one order of magnitude larger. Following Ormel et al. (2015), the Kelvin-Helmholtz timescale 
scales with the square of the gas-to-core ratio, we therefore obtain the following timescale in the case of dust-free gas:
\begin{equation}
t_{\rm KH} =  4 \times 10^6   \chi^2   \left( {Z \over 0.02} \right)^{0.25}  \left( { M_{\rm core} \over 5 M_\oplus } \right) ^{-3.93} {\rm yr}
\end{equation}
In the case of dusty atmosphere, the corresponding timescale is:
\begin{equation}
t_{\rm KH} =  4 \times 10^7   \chi^2   \left( {Z \over 0.02} \right)^{0.72}  \left( { M_{\rm core} \over 5 M_\oplus } \right) ^{-3.93} {\rm yr}
\end{equation}
Note that these timescales are valid for not too high values of Z (Lee et al. 2014 quote a value of the order of $Z \sim 0.5$), as for
higher values of Z, the increased mean molecular weight can decrease the KH timescale (see Hori and Ikoma, 2011).

{Finally, we emphasize the fact that in all this paper, we assume that the composition of the envelope is uniform. This is probably the case
if convection is vigorous enough in the envelope. However, if convection is not efficient enough, the polluted portions of the envelope (the innermost
parts at the beginning of the formation) may contract more rapidly, whereas the non-polluted parts (outermost regions of the envelope at the beginning
of formation) would contract slowly. This may result to the fact that the innermost region would resist more to the replenishment than the outermost regions.
The precise determination of the envelope contraction as well as of the accurate efficiency of the replenishment is beyond the scope of this paper
and will be the subject of future studies.}

The third important timescale is the accretion timescale of planets that is defined as 
\begin{equation}
t_{\rm acc} = { M_{\rm core} \over \dot M_{\rm solids}  } 
\end{equation}
The accretion rate of solids will be taken to be either $10^{-5} \mearth/$yr in the case of pebble accretion,
or $10^{-6} \mearth/$yr in the case of planetesimal accretion (see below).

Finally, the fourth important timescale is the pollution timescale, which quantifies on which timescale the
planetary envelope is  polluted when accreted solids are completely destroyed in the envelope:
\begin{equation}
t_{\rm pollution} = { \chi M_{\rm core} \over \dot M_{\rm solids}  } 
\end{equation}

\subsection{Destruction of accreted solids}
\label{destruction_solids}

When accreted solids enter the gas envelope of the planetary embryo, they suffer four mechanism:
\begin{itemize}
\item the gravitational interaction with the planet
\item the gas drag which depends on the thermodynamical properties in the envelope and the  characteristics
of solids (essentially the size and density)
\item thermal mass loss by melting or vaporization (Podolak et al. 1988, Lozovsky et al. 2017)
\item mechanical destruction when the pressure difference on the solid is larger than its internal strength. For the solids
we consider here, the self-gravity is not important, and the internal strength is only due to material tensile strength. 
\end{itemize}

Taking into account these four processes, Mordasini, Alibert and Benz (2006, MAB06) computed the mass of the envelope needed
to destroy an incoming stony body during a central impact before it reaches the core. {A detailed description of this model
can be found in Mordasini et al. (2015), where the main assumptions behind the model, as well as the major equations
solved in order to compute the solid-envelope interaction are presented.} As shown in MAB06, two
main effects lead to the destruction of planetesimals. For small bodies (below ~300m to ~1km) thermal ablation is the main process
leading to the destruction, whereas for larger bodies, mechanical destruction is the main effect. Interestingly enough, and as noted
in MAB06, these two effects lead to the fact that intermediate bodies (around 300m to 1 km in size)
are very resistant to the destruction and can reach a core surrounded by and envelope as big as $\sim 1-3 M_\oplus$. For the case
considered in this paper, the envelope mass needed to destroy solids of size $s$ before they reach the core is approximately given by
the following fit of the numerical results of MAB06:
\begin{itemize}
\item ${ M_{\rm enve} \over 0.001 M_\oplus} = \left( {s \over {\rm 10 cm}} \right)^{1.17}$ for $ s < 20 {\rm m}$
\item ${ M_{\rm enve} \over 0.5 M_\oplus} = \left( {s \over {\rm 20 m}} \right)^{0.66}$ for $20 {\rm m} < s < 300{\rm m}$
\item $M_{\rm enve} \sim 3 \mearth$ for $300 {\rm m} < s < 1.5 {\rm km}$
\item ${ M_{\rm enve} \over 0.03 M_\oplus} = \left( {s \over {\rm 1.5 km}} \right)^{1.06}$ for $ s > 1.5 {\rm km}$
\end{itemize}

The first and second regime corresponds to thermal destruction (ambiant gas and shock heating), the fourth regime
corresponds to mechanical destruction (sizes larger than 1.5 km). {In this latter case (largest planetesimals), self-gravity is important
and is included in our code (see Mordasini et al. 2015)}. Note that these fits corresponds to central impacts
(impact parameter equal to 0). For more general impacts (off-axis), the effect of gas drag is larger during the incoming
trajectory. As a consequence, the masses quoted above correspond to maximum values: if we consider all the possible impacts,
solids will be destroyed by smaller envelope masses.
 
 In the case of mechanical destruction, matter is released in the envelope as small particles. If the temperature in the envelope
 is not large enough, these small particles may sink towards the core. If the temperature is large enough (larger than $\sim 1600$K,
 corresponding to the sublimation of silicates, see Thiabaud et al. 2014), all the material of accreted solids is released as gas in the 
 planetary envelope and may be subject to the recycling back to the protoplanetary disk (see Sect.\ref{timescales}). This material will 
 in this situation not contribute to the mass growth of the planet. We plot in Fig. \ref{Tbase1600} the temperature at the base of the envelope, 
 for planets with an envelope equal to $10^{-4} M_\oplus$, for different epochs, and different locations. As can be seen on the plot, the
 temperature is always large enough to vaporize silicates if they are released as small dust particles. 
        
As we mention above, the luminosity of the core is computed by assuming that the accreted solids reach the core. This may seem
 \textit{a priori} contradictory with the fact that we are interested by the destruction of the solids during their travel towards the core. 
 However, up to the point where the envelope is massive enough to destroy the solids (the mass we precisely want to compute), this 
 assumption is justified. Moreover, the maximum core mass the planetary embryo can attain before solids are destroyed does not 
 depend very strongly on the value of the core luminosity. We discuss in more details this aspect in Sect. \ref{evolution}.

\section{Results}
\label{results}
\subsection{Envelope mass and timescales}

\begin{figure}
\hspace{-0.2cm} \includegraphics[height=0.30\textheight,width=0.40\textheight,angle=0]{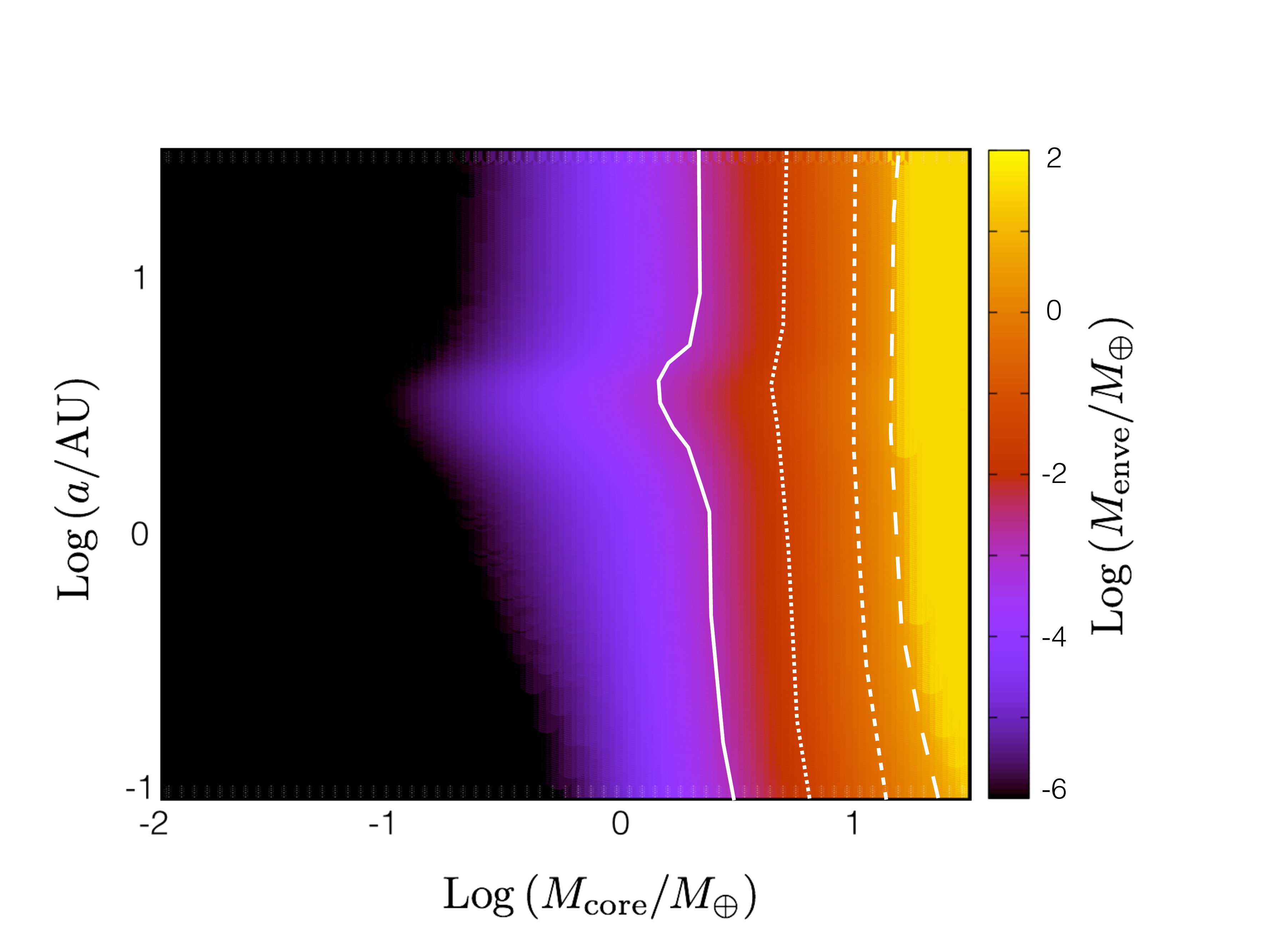}
  \caption{Mass of planetary envelope as a function of semi-major axis and core mass. The boundary conditions used for the internal structure
  calculations are the ones in the protoplanetary disc at 1 Myr, and the accretion rate of solids is equal to $10^{-6} M_\oplus /$yr.  {The contours represent
  the typical values quoted in Sect. \ref{destruction_solids}, namely, from left to right, $0.001 M_\oplus$, $0.03 M_\oplus$, $0.5 M_\oplus$ and $3 M_\oplus$} }
  \label{mass}
\end{figure}

\begin{figure}
\hspace{-0.2cm} \includegraphics[height=0.30\textheight,width=0.40\textheight,angle=0]{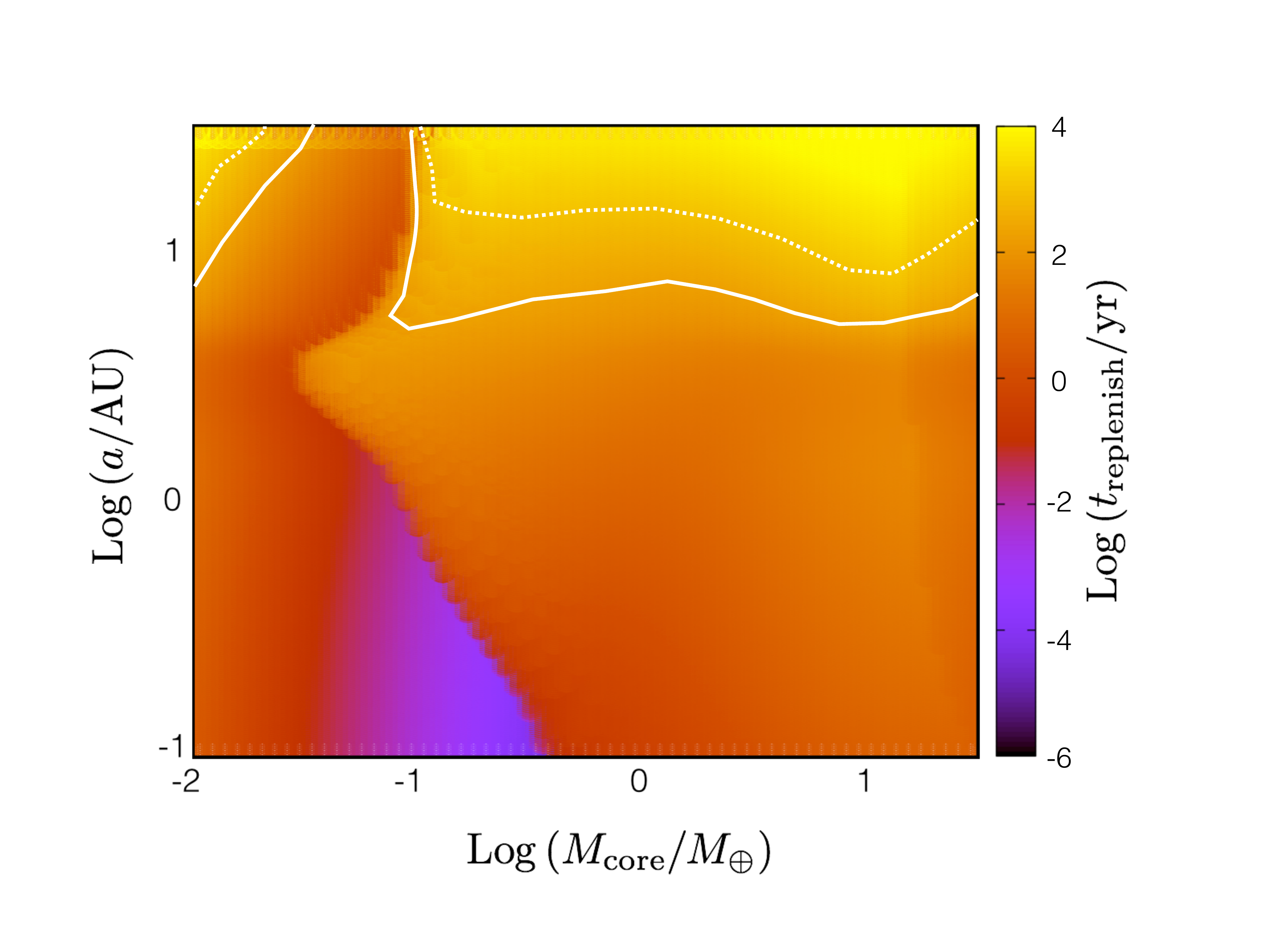}
  \caption{Replenishment timescale for the same planetary embryos as in Fig. \ref{mass}. {The contours represent
  the location where the timescale is equal to 200 years (solid line) and 1000 years (dotted line), values typical to the settling and growth timescale of dust according
  to Mordasini (2014)}. }
  \label{treplenish}
\end{figure}

We first consider an accretion rate of solids of $10^{-6} M_\oplus  /$yr, which is typical for planetesimal, and the state of
the protoplanetary disc at 1 Myr. Fig. \ref{mass} shows the mass of planetary envelopes as a function of core mass and distance
to the star, and Fig. \ref{treplenish} shows the replenishment timescales for the same planetary embryos. 

{As can be seen on Fig. \ref{mass}, the larger the core mass, the larger the envelope mass, as we expect as a result of the
increased core gravity. In addition, one can notice that the envelope mass, for a given core mass, increases for larger semi-major
axis. This is also something that is expected, as the gas entropy is lower at larger semi-major axis. The structure seen for a semi-major
axis equal to a few AU is due to the change of disk structure at the iceline. The replenishment timescale (Fig. \ref{treplenish}) also depends on the core mass and the semi-major axis. The structure
that can be seen on the figure result from the change of the envelope mass as a function of the core mass and semi-major axis,
as well as the disk properties that depend on the distance to the star (see Bitsch et al. 2015).}

In the case of larger accretion rates like the ones encountered during the accretion of pebbles ($\sim 10^{-5} \mearth/$yr, Lambrecht et al. 2014),
the results obtained are similar, the envelope mass and the replenishment timescale being slightly smaller.

\subsection{Maximum core mass}

\subsubsection{Accretion of pebbles}

\begin{figure}
\hspace{-0.2cm} \includegraphics[height=0.30\textheight,width=0.40\textheight,angle=0]{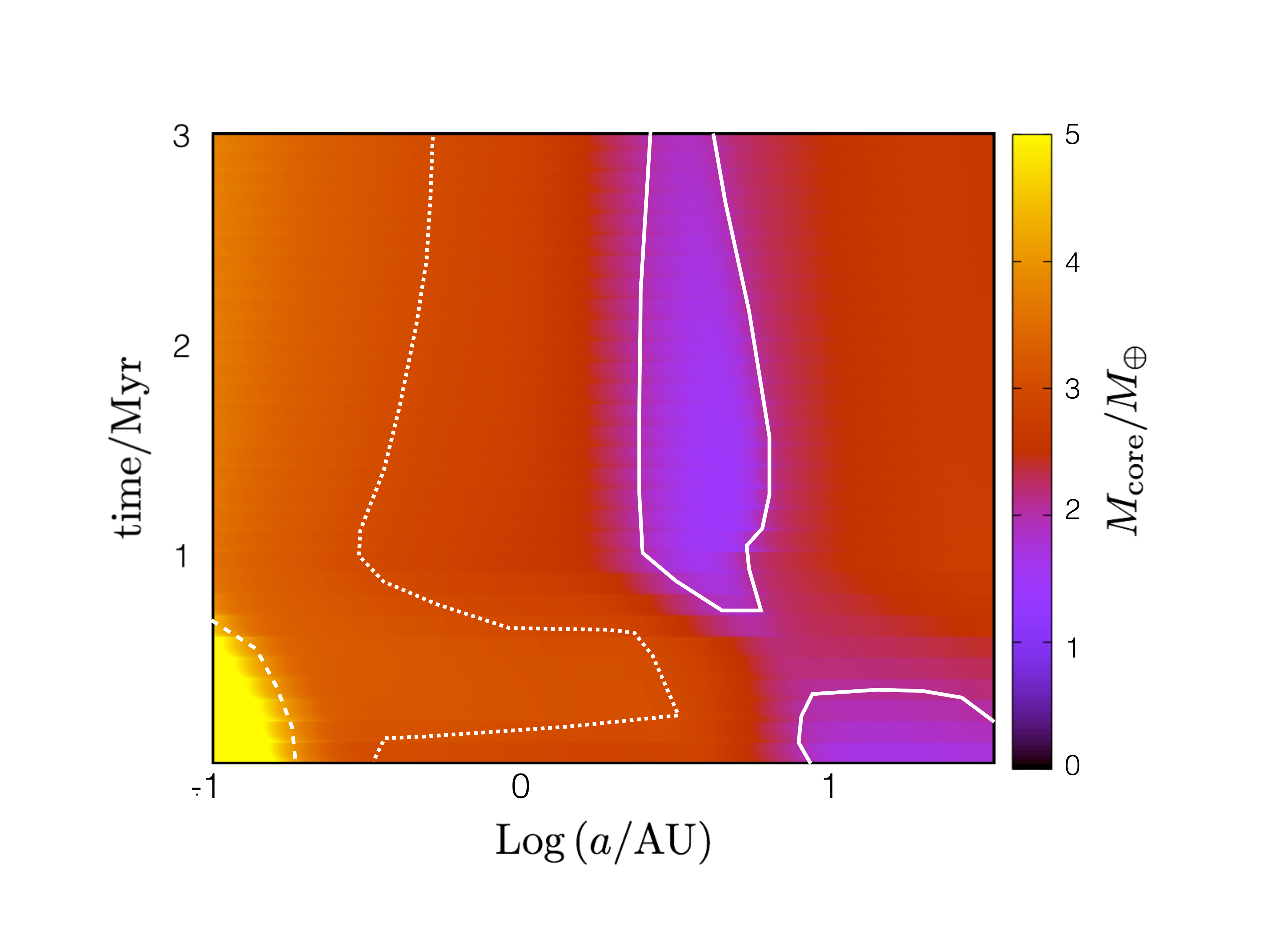}
  \caption{Core mass required to bind an envelope mass of 0.001 $\mearth$, for different semi-major axis,
  and at different epochs. {The contours represent core masses equal to $2 M_\oplus$ (solid line), $3 M_\oplus$ (dotted line) and $4 M_\oplus$ (dashed line)}}
  \label{Menve_0.001}
\end{figure}

\begin{figure}
\hspace{-0.2cm} \includegraphics[height=0.30\textheight,width=0.40\textheight,angle=0]{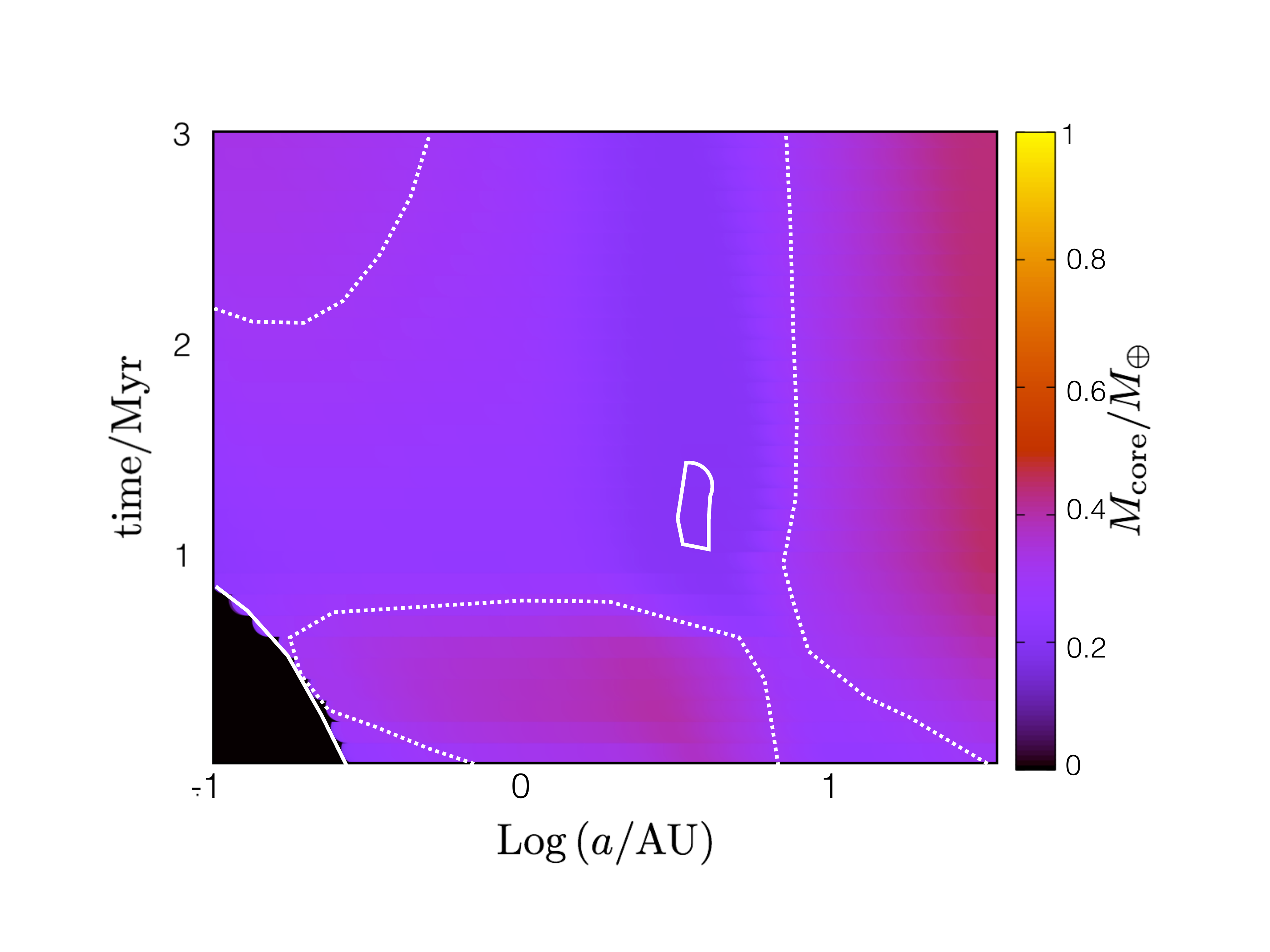}
  \caption{Core mass required to reach a temperature of 1600K at the base of the envelope, for different semi-major axis,
  and at different epochs. {The contour represents a core masse equal to $0.2 M_\oplus$ (solid line) and  $0.3 M_\oplus$ (dotted line). }}
  \label{Tbase1600}
\end{figure}

We now turn to the computation of the maximum core mass that is needed to bind an envelope large enough to destroy pebbles. If we first consider pebbles
as stony bodies of 10cm in size, an envelope mass of 0.001 $\mearth$ is large enough to destroy them. If we consider that pebbles are a mixture of icy and
stony grains, they are less resistant to high temperature, and we can assume that, as soon as the temperature at the base of the envelope is larger than
1600 K, they are destroyed in the envelope. Figures \ref{Menve_0.001} and \ref{Tbase1600} show the core mass that is required to match either the first
{(envelope mass large enough to destroy incoming solids)} or the second {(temperature at the base of the envelope
larger than 1600K)} criterion, for semi-major axis ranging from 0.1 AU to 30 AU, and at an epoch ranging from 0 to 3 Myr. For this calculation, the accretion rate 
of solids is equal to $10^{-5} \mearth/$yr, a value that is typical for pebble accretion (Lambrecht et al. 2014). Using the first criterion, the maximum mass
obtained is of the order  $\sim 1$ Earth masses, except very close to the star and at very early epochs. Using the second criterion, the maximum mass obtained
is a fraction of an Earth mass. We recall that the first criterion is derived from central impacts of stony bodies, which are much more resistant than are likely to be pebbles.
Therefore, it is likely that at masses lower than the one showed in Fig. \ref{Menve_0.001}, pebbles are vaporized in the envelope. Note finally that the temperature
at the base of the envelope in the case shown in Fig. \ref{Menve_0.001} is always larger than 1600K. In the case of radiative envelopes (as it is the case here),
these high temperature are the result of the integration of the diffusion equation for the radiative flux (see Stevenson 1982). As we mentioned above (Sect. \ref{envelope}),
these equations are valid for optically thick envelopes, themselves requiring a core mass larger than a fraction of an Earth mass. As showed in Fig. \ref{Menve_0.001},
the core mass for pebble destruction are of the order of $\sim 1 \mearth$, validating our approach.

Whatever the criterion used, when the mass of the protoplanetary core is larger than $\sim 0.5-3 \mearth$, pebbles are vaporized in the atmosphere. 
As a consequence, the growth of the protoplanetary core stops at a mass that is of the order of the mass of the Earth, the value that will be used in the following.
Any further growth must therefore result from either Kelvin-Helmholtz contraction (accretion of H/He in the envelope) or from the ability of the planet to retain 
vaporized heavy material (replenishment timescale larger than the accretion timescale). 

\subsubsection{Accretion of planetesimals}

\begin{figure}
\hspace{-0.8cm} \includegraphics[height=0.25\textheight,width=0.35\textheight,angle=0]{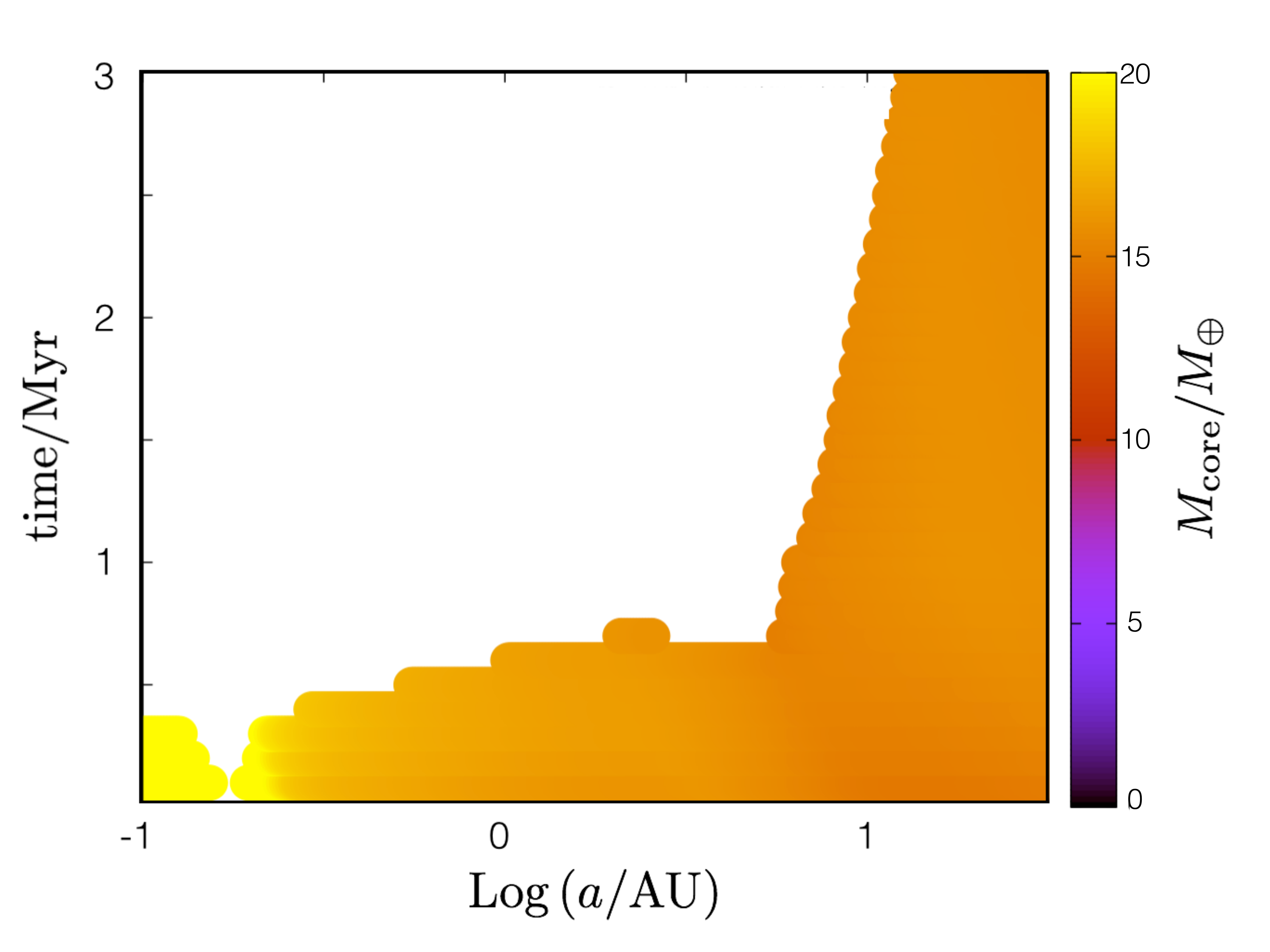}
  \caption{Core mass required to bind an envelope mass of 1 $\mearth$, for different semi-major axis,
  and at different epochs. The white regions of the plot correspond to planets that are not embedded in the disc ($H_{\rm disc} < R_{\rm Bondi}$).}
  \label{Menve_1}
\end{figure}

We now consider planetesimals of large size. In this case, the accretion rate used to compute the planetary envelope structure is equal
to $10^{-6} \mearth / $yr (e.g. Alibert et al. 2005), and an envelope mass of $1 \mearth$ is required to completely destroy a stony body of 40
km in size (MAB06), much more massive than in the case of pebbles (see previous section). The core mass that is required to bind such a massive
envelope is correspondingly more massive than in the case of pebbles, as shown in Fig. \ref{Menve_1}, and reaches more than 15 $\mearth$.
Note that, in large ranges of the parameter space (white regions in the figure), the disc scale-height is smaller than the Bondi radius of the planet,
and this latter is not embedded. In this case, it is not clear if the replenishment process actually works. If not, there is no limit to the growth of the
core due to the destruction of accreted solids.

\subsection{Evolution after the maximum core mass}
\label{evolution}
\subsubsection{Accretion of pebbles}

An important question that arises from our results is the fate of planets once the maximum core mass has been attained.
First, we must wonder if the process responsible of this maximum mass (the vaporization of pebbles) still holds after this point. Indeed,
when pebbles are destroyed in the envelope, they do not heat up anymore the base of the envelope, and the temperature at this location
could decrease, then allowing again pebbles to reach the core. However, the planets we consider here are radiative, and it is known
that in this case, the temperature at the base of the envelope depends on the mass of the planet, and the mean molecular weight of
the gas (Stevenson 1982):
\begin{equation}
T_{\rm base} = {G M_{\rm planet} \mu m_H \over 4 k_B R_{\rm core}} 
\end{equation}
where $R_{\rm core}$ is the radius of the planetary core. It is important to note that this temperature does no depend on the
luminosity of the core (which comes from the accretion of pebbles).

Once pebbles are destroyed in the envelope, the mean molecular weight increases, which leads to an increase of the
temperature at the base of the envelope. We can therefore conclude that, once pebbles start to be vaporized in the envelope,
the process does not stop, even if the pebbles do not deposit anymore their energy at the planetary core.

\begin{figure}
\includegraphics[height=0.25\textheight,angle=0]{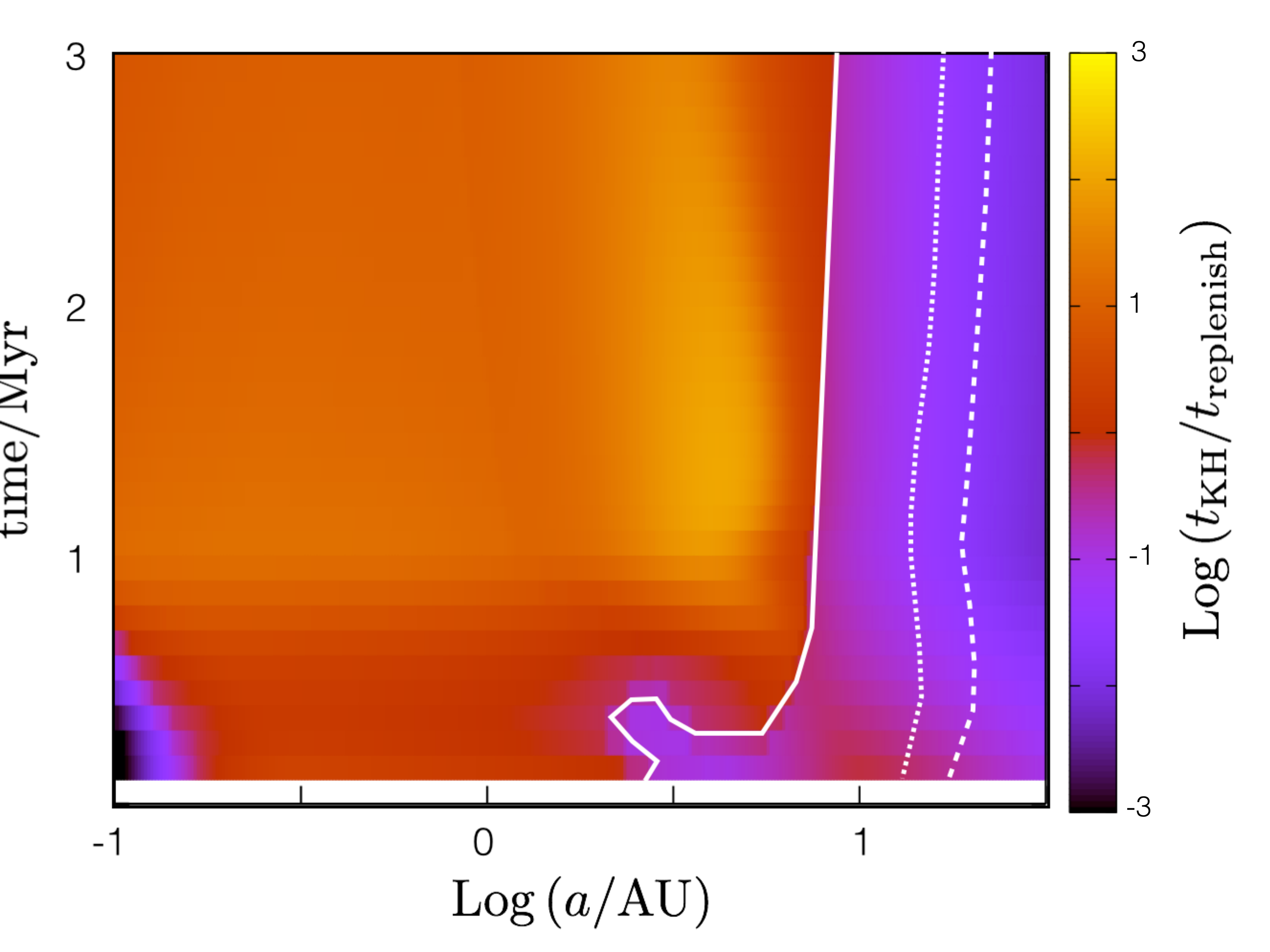}  
  \caption{Ratio of the Kelvin-Helmholtz to the replenishment timescales at the onset of pebble vaporization, as a function of time and
  semi-major axis. The white solid line shows the planets for which the replenishment and the pollution timescales are equal. Planets on the left side
  have a replenishment timescale smaller than the pollution timescale, meaning that their envelope has a composition similar to the one of
  the disk. Planets on the right side have a replenishment timescale larger than the pollution timescale, and the envelope is highly enriched
  in heavy elements. The white dotted line shows the planets for which the envelope enrichment is larger than 0.8, and the white dashed
  line the planets for which the envelope enrichment is larger than 0.9. }
  \label{ratio}
\end{figure}

Then, the energy supply at the base of the planet's envelope is suppressed, and the planet evolves by Kelvin-Helmholtz contraction.
The pollution time (see above) can however in certain case become comparable to  the replenishment time, which implies that the planetary envelope
may become heavily polluted rapidly. The ratio of the KH to the replenishment timescale is plotted in Fig. \ref{ratio} 
as a function of the location and the time. The white solid line shows the planets for which the pollution timescale is equal to the replenishment 
timescales. For planets on the left side (semi-major axis smaller than $\sim 10$ AU), the replenishment timescale is smaller than the 
pollution timescale. The material that is released from the vaporization of pebbles is therefore lost to the
protoplanetary disk before it can accumulate in the envelope. The total amount of heavy elements in the planets stops growing
and mass growth can only result from the accretion of H/He from the disk on a KH timescale. {The replenishment is
very fast, the timescale being smaller than $\sim$ 100 years in general (see Fig \ref{treplenish}). This timescale is smaller that the dust growth timescale (see Mordasini 2014),
and as a consequence,  the KH timescale has to be evaluated for dusty envelopes (the grains have the same size and composition
as in the ISM, following Lee et al. 2014). In addition,  and for the same reason, the KH timescale has to be evaluated}
for the metallicity of the disk (assumed to be equal to 0.02):
\begin{equation}
t_{\rm KH} =  4 \times 10^7   \chi^2  \left( { M_{\rm core} \over 5 M_\oplus } \right) ^{-3.93} {\rm yr}
\end{equation}

For planets located further out (right side of the white line on Fig. \ref{ratio}, semi major axis larger than $\sim 10$ AU), the replenishment timescale is longer than the pollution 
timescale, and the metallicity in the planet increases. Taking $Z=0.5$ as a typical value\footnote{The actual value of Z results from the competition between accretion 
and replenishment, and is given by $Z = { t_{\rm repl} \over t_{\rm repl} + t_{\rm poll}}$. The value of Z=0.5 therefore corresponds to the white solid line
on Fig. \ref{ratio}}, the Kelvin-Helmoltz timescale is given by:
\begin{equation}
t_{\rm KH} =  8.9 \times 10^6   \chi^2  \left( { M_{\rm core} \over 5 M_\oplus } \right) ^{-3.93} {\rm yr}
\end{equation}
for the dust-free case (it is more than one order of magnitude larger for dusty envelopes). 

\vspace{0.3cm}

As can be seen on Fig. \ref{ratio}, the KH timescale is larger than the replenishment timescale for nearly all planets below 10 AU (except
very close and at very early epochs). In this case, and as demonstrated in Ormel et al. (2015), the accretion of H/He gas cannot happen. As the core
cannot grow, and the metallicity of the planetary envelope is maintained at low values by the replenishment, the mass growth
of the planet is  stopped altogether.

For planets located at larger distance, the Kelvin-Helmholtz accretion occurs, increasing the amount of H/He in the planetary envelope. As the pollution
timescale is shorter than the replenishment timescale\footnote{Note that, as long as the envelope mass is negligible compared to the total planetary mass,
the pollution and the replenishment timescales scale with the envelope mass. The ratio between both timescales is therefore, as a first order approximation,
independent of the envelope mass.}, the accreted H/He gets rapidly polluted. 
{For example, on the right side of  lines showed in Fig, \ref{ratio}, the envelope enrichement is larger than 0.8 (white dotted line) or 0.9 (white dashed line).
In these regions, the KH timescale could become very small due to the increased mean molecular weight, triggering gas accretion, at least if dilution due to this latter
  does not decrease again the metallicity (Hori and Ikoma, 2011, Venturini et al., 2016).}

\begin{figure}
\includegraphics[height=0.25\textheight,angle=0]{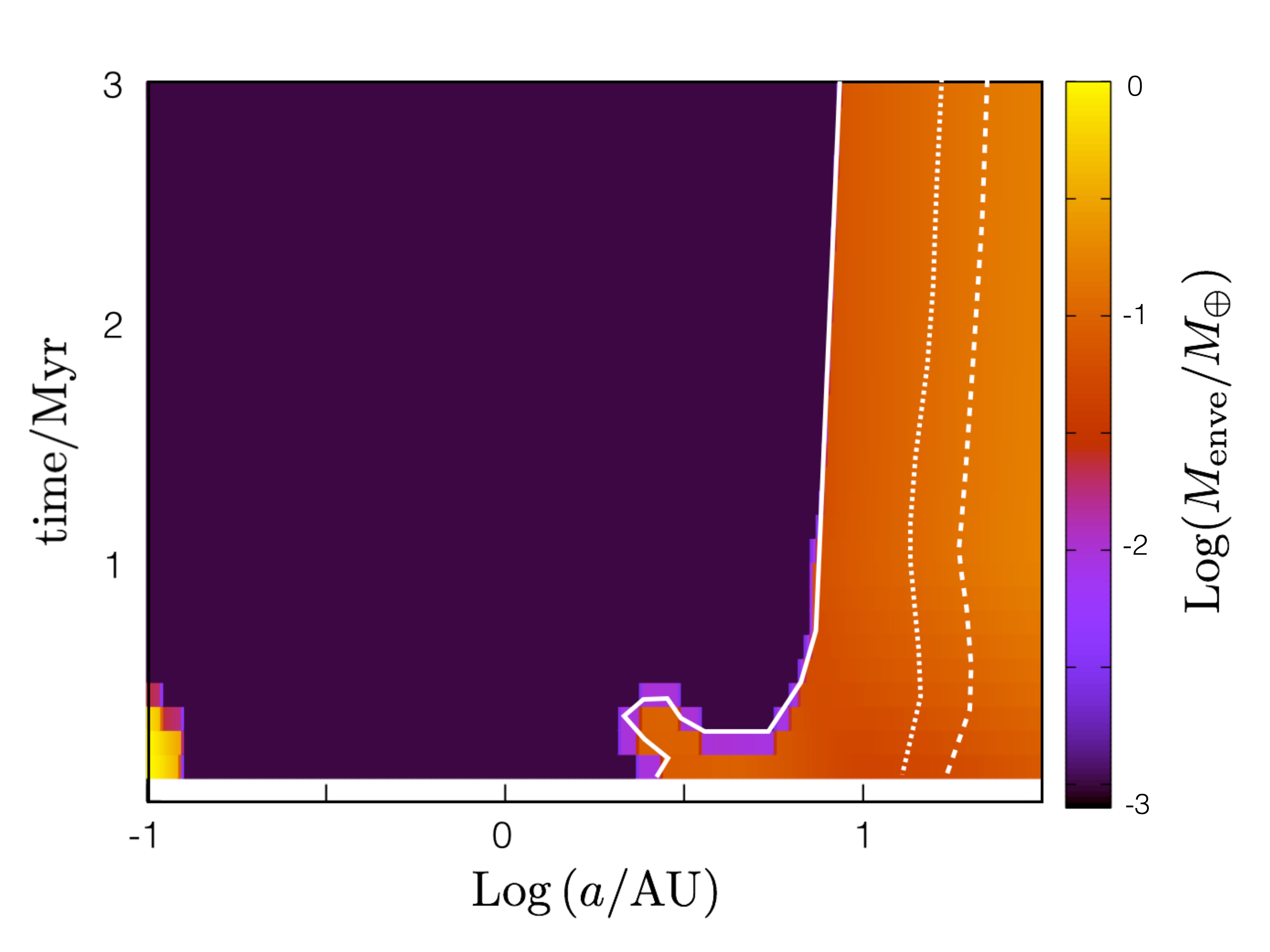} 
  \caption{Maximum envelope mass as a function of location in the disk and epoch. The white contours have the same meaning as in Fig. \ref{ratio}.}
  \label{menve_max}
\end{figure}

In these parts of the diagram, the planetary envelope then grows on the KH timescale,
the metallicity being kept large. {As stated in Ormel et al. (2015), and following the analytical estimates of Piso and Youdin (2014),
the KH timescale scales with the square of $\chi^2$ (the envelope to total mass ratio). This scaling results from the fact that the energy content of the envelope
scales with the pressure at the radiative-convective boundary, itself scaling with the envelope mass. In addition, the luminosity of the planet is inversely proportional
to this latter pressure. The KH timescale is therefore quadratic in the envelope mass, therefore in $\chi$. We emphasize the fact that this scaling results from the
simplified 2-layer model of Piso and Youdin (2014), and that numerical simulations are necessary in order to derive more precisely the evolution of a planet
once the core growth has stopped. Adopting this scaling for the rest of the paper, we see that, since the }
replenishment timescale scales with $\chi$, the two timescales become equal for some value of the envelope mass . Beyond this point, the replenishment prevents the cooling of the envelope,
and no more H/He accretion occurs (Ormel et al. 2015). As the core cannot grow anymore (solids are destroyed in the envelope), and the accretion
of H/He from the disk is prevented by the replenishment timescale, the growth of the planet stops altogether. The envelope mass that the planet can
accrete is plotted in Fig. \ref{menve_max}, as a function of location and time. As explained above, its value is equal to $10^{-3}$ for planets inside $\sim 10$ AU
(the KH timescale is larger than the replenishment timescale at the onset of pebble vaporization), and grows to a fraction of an Earth mass for
planets further away. Note that we have not considered here the time it would require to accrete this final envelope mass. If this time is long compared
to the disk lifetime, the final envelope would be even lower.
  
\vspace{0.3cm}

Finally, if the planet succeeds to reach a very high metallicity (of the order of 1) in the envelope, the KH timescale can be reduced substantially (Hori and Ikoma 2011).
For example, a one Earth mass core surrounded by an envelope of Z=0.8 metallicity has an accretion timescale of 1 Myr, which is comparable to the disk lifetime.
Any planet, whose envelope metallicity is larger than this threshold, could therefore accrete a substantial amount of H/He from the disk before the
disk has disappeared. This process, however, can only take place in the outer regions of the disk. 

The dotted and dashed white lines  on Figs. \ref{ratio} and \ref{menve_max}, give planets with  an envelope metallicity equal to  0.8
and 0.9 respectively. This corresponds in our nominal model to planets located at a distance larger than $\sim 20$ AU, but the location of these lines depends sensitively on the value
of the accretion rate of solids. For example, for an accretion rate of solids ten times smaller than our nominal value, the solid white line would be translated to the right,
and would be located close to the dashed white line.

In addition, as H/He accretion proceeds, the metallicity in the envelope can be reduced by dilution (Hori and Ikoma 2011), which could re-increase the KH timescale,
and therefore slow down accretion. Taking into account these effects requires computing the planet evolution in a way similar to Venturini et al. (2016), and is
beyond the scope of this paper.

{The result and the discussion we have presented in this section depend on the value of the Kelvin-Helmholtz timescale. Interestingly enough, the value of the KH timescale
derived by Hori and Ikoma (2010) appear smaller than the ones of Lee et al. (2014) that we use in this paper. If the KH timescale is reduced compare to the values assumed here, 
planets could acquire a large envelope at distance smaller that the ones presented in Fig. \ref{menve_max}.}

\subsubsection{Accretion of planetesimals}

In the case of the accretion of planetesimals, we can make the same estimations. However, as the limiting core mass
is much larger, the KH timescale is much shorter. As a consequence, once the maximum core mass is reached, the planet
will start accreting gas and, depending on the remaining lifetime of the disk, become a Neptune-like planet or a gas giant.

Interestingly enough, for an envelope mass of $\sim 1 \mearth$ and an accretion rate of $10^{-6} \mearth / $yr, the pollution timescale
is much larger than the replenishment timescales. As a consequence, the heavy elements released by the destruction of solids cannot
accumulate in the envelope, whose metallicity remains small. We therefore expect, in this scenario, that the planetary envelope
is made of nearly pure H/He (the metallicity being the one of the gas in the protoplanetary disk), {at least as long as the replenishment
is efficient. We note that some planets are likely to contain a very large mass of heavy elements (e.g. HD149026b, see Ikoma et al.  2006, Guillot
et al., 2006).  The formation of such a planet by accretion of planetesimals is not hindered by the replenishment process, provided
this large mass of planetesimals is accreted after the replenishment has ceased, when the planet Hill radius is larger than the disk scale height.} 

 \section{Discussion}
 \label{discussion}

  The maximum core mass a planetary embryo can reach before its envelope is so large that accreted solids are vaporized depends
  strongly on the size of these latter. In the case of pebbles of $\sim 10$ cm in size, the maximum core mass is of the order of
  one Earth mass. In the case of planetesimals a few kilometers in size, the maximum core mass is larger than $\sim 15 \mearth$.
  This results have strong implications: if the replenishment timescale is shorter than the accretion timescale, as shown by Ormel et al. (2015) 
  in their simulations, the material vaporized in the planetary envelope is lost on a timescale that is shorter than the timescale on which
  solids are accreted. This means that the material originating in accreted solids does not accumulate in the planet, and cannot contribute
  to its mass growth. The interplay between the disruption of solids due to their interaction with the gas envelope, and the strong advection
  wind originating from the protoplanetary disc, therefore leads to the end of planetary growth for a core mass that depends on the size
  of accreted bodies. In the case of pebble accretion, the growth stops around the mass of the Earth or at smaller masses, for the case
  of planetesimals, the growth stops at a mass larger than 15 $\mearth$. As we demonstrated above, further growth by accretion of H/He 
  from the protoplanetary disk is negligible in the case planets are small (Earth mass). In this case, the growth of planetary embryos is stopped
  by the process of solids destruction in the envelope. Any further planetary growth must proceed  by collision between planetary embryos,
  or must wait until the thermodynamical conditions have changed in the protoplanetary disk.  In the case accreted solids are much bigger, 
  and the core needed to bind an envelope large enough to destroy them is also large, the planetary growth is not stopped, as accretion of 
  H/He from the protoplanetary disk is allowed by the short KH timescale.
  
  In the case of pebble accretion, the maximum core mass is very small, and the accretion of gas once the core growth has ceased is
  very small, and no noticeable accretion of gas can proceed during typical disk lifetimes. As a consequence, the growth of planetary embryos
  by pebble accretion is not possible for masses beyond $\sim 1 \mearth$, at least in the innermost regions of the disk (semi-major axis 
  smaller than $\sim 20$ AU). The formation of, for example, Jupiter and Saturn in the innermost 20 AU of the disk by pebble accretion
  would require either that they form by the collision between bodies of $\sim 1 \mearth$, or that some of the assumptions used in this paper are
  not fulfilled. The first hypothesis would mean that the planetary embryos collide on a short timescale, because a mass of $\sim 10 \mearth$
  must be reached before the gas disk has disappeared, in order for the planet to have enough time to accrete gas. This poses a problem,
  as, when the protoplanetary disk is present, planetary embryos should be kept on quasi-circular orbits as a result of disk-embryo interactions.
  In this case, one expect that substantial collision would occur only when the disc has nearly disappeared, at a time when there is probably
  not enough gas to form the envelope of Jupiter and Saturn. The formation of Jupiter and Saturn, in this model, should therefore happen at
  distances larger than $\sim 20$ AU, followed by, or simultaneously with, a phase of migration.
  
Another possibility is that some of the assumptions used
  in this work are not fulfilled during the formation of planets. For example, the calculations performed by Ormel et al. (2015) are based on isothermal
  equation of state and we have assumed a value of $f_{\rm cover}^*$ that does not depend on the planetary mass nor on the semi-major axis
  (we have adopted a value that is on the lower end of the range derived by Ormel et al. (2015), the replenishment timescale we obtain
  are therefore upper limits). If more realistic models would show that the replenishment timescale is much longer (longer than the accretion timescale),
  the replenishment of planetary envelopes would be negligible. In this case, the growth of planets once pebbles are disrupted in the core
  could continue by simultaneously growing the core, and increasing the metallicity of the envelope (see Venturini et al. 2015, 2016). 
  
  {We also note that the simulations performed by Ormel et al. (2015) assume an isothermal and inviscid gas. On the other hand, other simulations e.g. by D'Angelo \& Bodenheimer (2013),
  including the effect of radiation transport and viscosity, found that material from the deep regions of the envelope are gravitationallly bound to the planet. The two simulations predict
different efficiency of the replenishment, as well as differences in the region of the envelope that can actually be replenished. If the material from incoming solids
  is dissolved in the innermost regions of the planet, and if this region is not replenished as showed in D'Angelo \& Bodenheimer (2013), the core growth would continue until a larger
  mass is reached. Indeed, as the core grows, the envelope mass also grows, and incoming solids are destroyed at higher an higher envelope levels. The termination of core growth would,
  in this case, occur at larger core masses.}
      
   \section{Conclusion}
 \label{conclusion}

 Using the results of MAB06, we have derived an approximate fit giving the envelope mass that is necessary to destroy
  stony solids of different mass before they reach the core of a forming protoplanet. The results of MAB06 are derived under the
  assumption of non-porous stony material, during a central impact. As impacts are in general non central, and solids are probably
  porous and/or made of mixture of silicates and ices (this is specially the case for pebbles that drift from the outer part of the protoplanetary
  disk, see Bitsch et al. 2015b), the envelope mass obtained by MAB06 are upper limit of the envelope mass. In other term, under more
  realistic conditions, the envelope mass that is needed to disrupt and vaporize accreted solids is probably smaller than we one we have used.
  As a consequence, since, for a given location in the disk, the envelope mass is a growing function of the core mass, the maximum core mass
  we derived in the previous section are probably over-estimated.

{We have shown that because of the interplay between the destruction of solids in the protoplanetary envelope and the replenishment
process, the core growth can be stopped at a mass that depends strongly on the typical size of accreted solids. For pebble accretion,
this size is of the order of the mass of the Earth, whereas, in the case of massive planetesimals (hundreds of meters at least), the limiting
mass is at least ten times larger. Once the core growth is stopped, any further growth must be the result of gas accretion which depends
on the ability of the planet to cool down. We have discussed this possibility using arguments based on the KH timescale, but
definitive conclusion will have to wait until the development of new formation calculations taking into account in a self consistent way: 1- solid destruction,  2- the consecutive enrichment
in heavy elements, and 3- the replenishment process.}

  Another conclusion of our work is that, in the case of planetesimal accretion, the pollution timescale of the planetary envelope is
  much longer than the replenishment timescale. This imply that, as long as the process of replenishment is active, the gas envelope 
  remains of low metallicity. Interestingly, the envelope of all the giant planets we know is enriched in heavy elements, this enrichment
  being very strong for Uranus and Neptune (e.g. Helled et al. 2011). In the framework of the models presented here, this imply that 
  a substantial fraction of solids are accreted once the replenishment process has ceased, for example when the disk scale-height
  becomes smaller than the Bondi radius of the planet (white regions on Fig. \ref{Menve_1}). We note that this is very  likely
  for planets forming at large distance from the star, as, in planetesimal-based planet formation, the accretion rate of solids is rather 
  slow. It is therefore likely that large fraction of the heavy elements is accreted at a stage when the replenishment 
  of the envelope is ineffective.
  
   Finally, we point out that the results presented in this paper are to be taken as proof-of-principle of the interplay between the advection
  wind and the vaporization of accreted solids in forming planets.  For example, one assumption of the calculations presented here is 
  that the growth of the core stops completely when solids are destroyed in the envelope. If some of the material would however manage 
  to reach the core, the picture would be changed, the efficiency of core growth being reduced and not suppressed \footnote{This could be for example
  the case if silicates would form droplets in the planetary envelope and if silicate would not be miscible enough in H/He gas at high pressure}.
  Moreover, the actual maximum core mass that a planet can reach in any given formation
  scenario depends on the precise internal properties of accreted solids (porosity, tensile strength, composition, size), as well as the dynamics
  of their accretion and then behavior of the material constituting the accreted solids at high temperature and pressure. Indeed, these factors govern 
  the solid-gas interaction in the planetary envelope, and ultimately the release of accreted material as gaseous species. Finally, we have
  not considered in this work the possibility that, once large enough metallicities are attained in the planetary envelope, solids may condense
  fast enough to be able to sink to the core. The computation of this effect requires the determination of the kinetics of condensation,
  of sinking as well as the knowledge of the thermodynamical properties of highly enriched material at high pressure, and is beyond
  the scope of this paper.  

{Note finally that Levison et al. (2015) presented a scenario for the formation of the solar system based on pebble accretion.
This scenario, which is specific to the formation of our system, seems to fit many of its dynamical constraints. However, this model did not include the replenishment
process of Ormel et al. (2015), and should, as a consequence, be revisited taking into account the possible effects described here.}

  Despite the shortcomings outlined above, the destruction of solids during the growth of planets, a process that is specially important in the case of
  pebble accretion, coupled with the replenishment of planetary envelopes, has strong implications on the growth of planets.
  Therefore, if the assumptions made in this work indeed hold (e.g. on the efficiency of replenishment), the process described 
  in this paper represents a serious bottleneck in the formation of planets more massive than a few Earth masses by pebble accretion,
  in particular in the innermost regions (below $\sim 10$ AU) of the disk.
            
\acknowledgements

We thank Willy Benz for insightful discussions. This work was supported in part by the European Research Council under grant 239605. 
This work has been carried out within the frame of the National Centre for Competence in Research PlanetS 
supported by the Swiss National Science Foundation. The authors acknowledge the financial support of the SNSF.


\begin{thebibliography}{}

\bibitem[2005]{AAbig} Alibert, Y., Mordasini, C., Benz, W., \& Winisdoerfer, C. 2005, \aap, 434, 343

\bibitem[2005a]{alibertetal2013} Alibert, Y., et al. 2013, \aap, 558, 109

\bibitem[]{BL} Bell, K. R. \& Lin, D. N. C., 1994, ApJ, 427, 987

\bibitem[2014]{benz14} Benz, W., Ida, S., Alibert, Y., Lin, D., \& Mordasini, C.,  2014, Protostars and Planets VI , 691

\bibitem[2015]{B15} Bitsch, B., Johansen, A., Lambrechts, M. \& Morbidelli, A., 2015a, 575, 28
 
 \bibitem[2015]{B15} Bitsch, B., Lambrechts, M. \& Johansen, A., 2015b, 582, 112
 
 \bibitem[2013]{dangelo13} D'Angelo, G. \& Bodenheimer, P., 2013, \apj, 778, 77
 
 \bibitem[2013]{fortier13} Fortier, A., Alibert, Y., Carron, F., Benz, W., \& Dittkrist, K.M., 2013, \aap, 549, 44
 
 \bibitem[2006]{guillot06} Guillot, T., et al. 2006, \aap, 453, L21
 
 \bibitem[1998]{hartmann} Hartmann, L., Calvet, N., Gullbring, E. \& D'Alessio, P. 1998, \apj, 495, 385

\bibitem[]{helled} Helled, R., Anderson, J. D., Podolak, M. \& Schubert, G., 2011, \apj, 726. 15

\bibitem[]{HI} Hori, Y. \& Ikoma, M., 2011, \mnras, 416, 1419

\bibitem[]{hubickyj} Hubickyj., O., Bodenheimer, P., \& Lissauer, J. J. 2005, Icarus, 179, 415

\bibitem[2004]{IL04} Ida, S. \& Lin, D.N.C. 2004, \apj, 604, 388

\bibitem[2000]{ikoma00} Ikoma, M., Nakazawa, K., \& Emori, H. 2000, \apj, 537, 1013

\bibitem[2006]{ikoma06} Ikoma, M., Guillot, T., Genda, H., Tanigawa, T. \& Ida, S. 2006, \aap, 650, 1150

\bibitem[2003]{Inaba03} Inaba,  S. \& Ikoma, M. 2003, \aap, 410, 711

\bibitem[2003]{inabaetal03} Inaba, S., Wetherill, G. W., \& Ikoma, M. 2003, Icarus, 166, 46

\bibitem[2010]{kobayashi10} Kobayashi, H., Tanaka, H., Krivov, A. V. \& Inaba, S. 2010, Icarus, 209, 836

\bibitem[2011]{kobayahsi11} Kobayashi, H., Tanaka, H., \& Krivov, A. V., 2011, \apj, 738, 35

\bibitem[2012]{LJ12} Lambrechts, M. \& Johansen, A. 2012, \aap, 544, A32

 \bibitem[2014]{L14} Lambrechts, M. \& Johansen, A. 2014, \aap, 572, 107
 
 \bibitem[]{lee14} Lee, E. J, Chiang, E. \& Ormel, C. W., 2014, \apj, 797, 95
 
 \bibitem[]{levison15} Levison, H, Kretke, K \& Duncan, M., 2015, Nature, 524, 322
 
 \bibitem[]{lissauer} Lissauer, J.J., Hubickyj, O., D'Angelo, G. \& Bodenheimer, P, 2009, Icarus, 199, 338
 
 \bibitem[]{michael} Lozovsky, M., Helled, R., Rosenberg, E. \& Bodenheimer, P., 2017, \apj, 836, 227

 \bibitem[2006]{MAB06} Mordasini, C., Alibert, Y.,  \& Benz, W., 2006,  in "Tenth Anniversary of 51 Peg-b: Status of and prospects for hot Jupiter studies. Colloquium held at Observatoire de Haute Provence, France, August 22-25, 2005.
  Edited by L. Arnold, F. Bouchy and C. Moutou
  
  \bibitem[2014]{M14} Mordasini, C., 2014, \aap, 566, 141
  
 \bibitem[2015]{M15} Mordasini,  C., et al., 2015,  Int. J. of Astrobiology, 14, 201

\bibitem[2010]{OK10} Ormel, C. W. \& Klahr, H. H. 2010, \aap, 520, A43

\bibitem[2016]{ormel2016}  Ormel, C. W., Shi, J.-M. \& Kuiper, R., 2015, MNRAS, 447, 3512

\bibitem[2014]{PY14} Piso, A.-M. \& Youdin, A., 2014, \apj, 786, 21

 \bibitem[1988]{Podolak88} Podolak, M., Pollack, J.B. \& Reynolds, R.T., 1988, Icarus 73, 163

\bibitem[1996]{Pollack96} Pollack, J. et al. 1996, Icarus, 124, 62

\bibitem[scvh]{scvh} Saumon, D., Chabrier, G. \& Van Horn, H. M. 1995, \apjs, 99, 713

\bibitem[]{stevenson} Stevenson, D. 1982, PSS, 30, 755

\bibitem[]{Thiabaud} Thiabaud, A., Marboeuf, U., Alibert, Y., Cabral, N., Leya, I. \& Mezger, K., 2014, \aap, 562, 27
  
 \bibitem[2015]{V15} Venturini, J., Alibert, Y., Benz, W. \& Ikoma, M., 2015, \aap, 576, A114
 
 \bibitem[2016]{V16} Venturini, J., Alibert, Y. \& Benz, W., 2016, \aap, \textit{in press}
   
\end{thebibliography}
\end{document}